%% file: main.tex
\documentclass[a4paper,11pt]{article}
\usepackage{aaskaiid}
\usepackage{orcidlink}
\title{Radio Galaxies and Jet Duty Cycles}
\ShortTitle{Radio Galaxies}

\author[1]{M. J. Hardcastle\orcidlink{0000-0003-4223-1117}}
\ShortName{M. J. Hardcastle et al.} % shortened name list for header 
\emailAdd{m.j.hardcastle@herts.ac.uk}
\author[2]{C. Konar\orcidlink{0000-0002-2530-3812}}
\emailAdd{chiranjib.konar@gmail.com}
\author[3]{S. Dutta\orcidlink{0000-0002-6542-2939}}
\emailAdd{sushant.dutta@uct.ac.za}
\author[4]{D. V. Lal\orcidlink{0000-0001-5470-305X}}
\emailAdd{dharam@ncra.tifr.res.in}
\author[5]{R. D. Baldi\orcidlink{0000-0002-1824-0411}}
\emailAdd{ranieri.baldi@inaf.it}
\author[5]{M. Brienza\orcidlink{0000-0003-4120-9970}}
\emailAdd{marisa.brienza@inaf.it}
\author[6]{A.~Hota\orcidlink{0000-0002-4959-7376}}
\emailAdd{hotaananda@gmail.com}
\author[7]{M. Kunert-Bajraszewska\orcidlink{0000-0002-6741-9856}}
\emailAdd{magda@astro.umk.pl}
\author[1]{B. Mingo\orcidlink{0000-0001-5649-938X}}
\emailAdd{b.mingo@herts.ac.uk}
\author[8]{M. Pandey-Pommier\orcidlink{0000-0001-5829-1099}}
\emailAdd{mamtapommier@gmail.com}
\author[9]{K.~Rubinur\orcidlink{0000-0001-5574-5104}}
\emailAdd{rubinur.khatun@astro.uio.no}
\author[10]{S.~Shabala\orcidlink{0000-0001-5064-0493}}
\emailAdd{stanislav.shabala@utas.edu.au}
\author[11]{E. Vardoulaki\orcidlink{0000-0002-4437-1773}}
\emailAdd{elenivard@gmail.com}

\affiliation[1]{Centre for Astrophysics Research, Department of Physics, Astronomy and Mathematics,
  University of Hertfordshire, College Lane, Hatfield AL10 9AB, UK}
\affiliation[2]{Department of Physics, Amity Institute of Applied Sciences, Amity University Uttar Pradesh, Sector-125, Noida 201313, U.P., India} 
\affiliation[3]{The Inter-University Institute for Data Intensive Astronomy (IDIA), and University of Cape Town, Private Bag X3, Rondebosch, Cape Town 7701, South Africa}
\affiliation[4]{National Centre for Radio Astrophysics - Tata Institute of Fundamental Research Post Box 3, Ganeshkhind P.O., Pune 411007, India}
\affiliation[5]{INAF - Istituto di Radioastronomia, Via P. Gobetti
  101, I-40129, Bologna, Italy}
\affiliation[6]{UM-DAE Centre for Excellence in Basic Sciences, University of Mumbai, Santacruz-East, Mumbai 400098, India}
\affiliation[7]{Institute of Astronomy, Faculty of Physics, Astronomy and Informatics, NCU, Grudzi\k{a}dzka 5/7, 87-100, Toru\'n, Poland}
\affiliation[8]{Scientific Centre, Catholic University of Lyon, 10
  Place des Archives 69288, Lyon, France}
\affiliation[9]{Institute of Theoretical Astrophysics, University of Oslo,
  Postboks 1029, Blindern, 0315 Oslo, Norway}
\affiliation[10]{School of Natural Sciences, University of Tasmania, Private Bag 37, Hobart 7001, Australia}
\affiliation[11]{National Observatory Athens, Hill of the Nymphs, Athens, Greece}

\abstract{Radio-luminous active galactic nuclei, or radio galaxies,
  are the brightest population of objects in the extragalactic radio
  sky and will be seen in large numbers in essentially every SKA
  observation. Despite having been studied for more than seventy
  years, some aspects of radio galaxy physics are still poorly
  understood, and the SKA will shed light on this by enabling the
  generation of very large samples of high-resolution, sensitive,
  broad-band images of radio galaxies, allowing us to probe, for
  example, regions of particle acceleration, spectral ageing, and the
  magnetic field structures both internal and external to the radio
  lobes. A key feature of the radio galaxy population is that
  observations of extended sources probe the past history, and thus
  the duty cycles, of accretion onto the central supermassive black
  hole, and we discuss ways in which the SKA will improve our
  understanding of episodic and dying radio galaxies in particular.}

\begin{document}
\include{journal-names}

\maketitle

\section{Introduction}
\label{sec:intro}

Radio galaxies and radio-loud quasars, sometimes called radio-luminous
active galactic nuclei (RLAGN) but referred to in this chapter simply
as radio galaxies, are a type of active galactic nucleus (AGN)
characterised by the fact that some or all of the nuclear activity
emerges in the form of a high-speed collimated jet of energetic
particles and magnetic fields. The synchrotron radio emission generated by these jets
makes the galaxies hosting such AGN particularly luminous in the
radio, giving them their name, but in fact emission from the jets and
the large-scale structures they generate can be seen across the
electromagnetic spectrum \citep[e.g.][]{Burbidge56}. Radio galaxies make up almost all of the
brightest sources seen in extragalactic surveys, with the transition
from radio galaxies to normal galaxies like the Milky Way (where radio
emission is driven primarily by star-formation activity) occurring at
the mJy level \citep{Condon+02}. Given the SKA's design, radio galaxies will be the
brightest population of discrete sources in almost every continuum and
polarization observation, whether they are the intended targets or
not.

The study of radio galaxies is distinct from that of other types of
active galaxies in several important ways. Firstly, the radio galaxy
population offers our only view of the distant AGN population below
the critical accretion rate (a few per cent of Eddington) at which
accretion ceases to be radiatively efficient
\citep{Hardcastle+07,Best+Heckman12,Mingo+14,Hardcastle18b}. While
some radio AGN have all the apparatus of a radiatively efficient AGN,
with a big blue bump, infrared-emitting torus, and X-ray corona, these
radiatively efficient objects, known as radio-loud quasars or
high-excitation radio galaxies (HERGs) and generally thought to
accrete through a standard thin disk \citep{Shakura&Sunyaev1973},
are in the minority in the radio galaxy population. Much more
common are the radiatively inefficient,
low-excitation objects (LERGs) which are thought to accrete via an
advection-dominated flow \citep[e.g.][]{Narayan&Yi1995}. The low
fraction of radiatively efficient objects is not surprising, since at any
given time, far more of the supermassive black holes at the centres of
galaxies are likely to be accreting at low rates than at high ones:
the radio galaxy population gives almost the only way of studying this
mode of accretion, sometimes referred to as low-luminosity AGN
(LLAGN).  However, going from radio observations to inferred 
accretion rates and thereby constraining the growth of these low-accretion 
rate black holes remains challenging. Jet generation almost certainly
involves quantities such as the spin of the parent black hole and the
magnetization of the accreting material \citep{Blandford+Znajek77}
which are not directly constrained in most cases by observation at any waveband.

Secondly, radio galaxies can grow to large sizes (up to several Mpc:
e.g. \citealt{Mostert+24})
in the course of their sometimes very long (up to $\sim 1$ Gyr) active
lifetimes, and their expansion through the interstellar and
intergalactic media and potentially into intergalactic voids
\citep{Oei+24} allows the transport of energy, cosmic rays and
magnetic fields from scales of the central black hole out to these
very large distances. Particularly of interest to the galaxy formation
and evolution community is their ability to heat the intracluster
medium and so, through a feedback loop, prevent the cooling of hot gas
and the subsequent formation of stars, thus playing a role in the
truncation of the galaxy mass function at the high-mass end
\citep[e.g.][]{Fabian12,McNamara+Nulsen12,Hardcastle+Croston20,Magliocchetti22,Eckert+24}.

Thirdly, and this is the most important point for the remainder of the
present chapter, radio galaxies give us a unique view of the
\textit{history} of jet (and therefore, presumably, accretion)
activity over the course of their lifetimes. Because the lobes of
radio galaxies grow with time, it has long been realised
\citep[e.g.][]{Baldwin82} that they evolve along tracks in the
power-linear size plot (the so-called $PD$ diagram) which is in some
sense analoguous to the Hertzsprung-Russell diagram for stars, and
that these tracks can be predicted given some model of the evolution
of the jets and their environment, for constant jet power (e.g.
\citealt{Kaiser+97}; see \cite{Turner+Shabala23} for a review). It
follows that distributions of source physical sizes can be used, with
a model, to infer information about the distribution of active
lifetimes of the population \citep[e.g.][]{Hardcastle+19}. In
addition, observations of radio galaxies with evidence for multiple
epochs of activity, so-called double-double radio galaxies
(i.e. radio galaxies with a double pair of lobes with a 
single core, see Section~\ref{sec:erg} for details) or episodic radio
galaxies \citep{Schoenmakers+00} have given one of the few available
pieces of
information on the duty cycle of AGN activity: these objects show, in
particular, that it is possible for the jet output to be initially
high, drop to a low level, and then return to something like the
original level on timescales that are short compared to the overall
radio galaxy lifetime, presumably as a result of changes in accretion
rate. In principle we can also detect the signatures of slower
variations in jet power, for example a slow decrease over a source's
lifetime, by looking for discrepancies betweeen the properties of the
current energy dissipation regions (jets and hotspots) and the
integrated properties of the lobes. In addition, the detection and
characterization of remnant radio galaxies, where jet activity has
completely ceased, is of great interest both because the evolution of
these objects is very sensitive to the physics of synchrotron loss and
expansion \citep{Godfrey+17,Hardcastle18,Turner+Shabala20}, and thus gives an important
probe of the overall AGN duty cycle \citep{Brienza+17,Shabala+20}, and
because the formation of a remnant is likely the precursor to mixing
of the lobe cosmic rays through the intergalactic medium
\citep{Brienza+21,Lalremnant2021}. Finally, observations of resolved
radio source morphology can give insights into the history of jet
direction and therefore (presumably) black hole spin orientation,
tracing both rapid changes in the jet axis due to black hole/black
hole merger \citep{Merritt+Ekers02} and slow reorientation due to
precession \citep[e.g.][]{Krause+19,Horton+20,Sebastian+24}. A key
question generally is how we can extract this information about
accretion and jet orientation history from the radio observations.
   
In this chapter we will review the advances that will be made in
understanding radio galaxy physics and life cycles using the completed
phase 1 SKA (AA4), focusing particularly on the population of AGN that
will be \textit{resolved} with the SKA. We explore how we can (i)
investigate the physical mechanisms driving jet production, (ii)
constrain theoretical models that aim to explain the diversity of
radio morphologies, (iii) find observational evidence for episodic jet
activity, and (iv) constrain ages of radio-plasma by spectral ageing
modelling to infer the duty cycle of episodic radio galaxies. We
assess the feasibility of addressing the key questions by presenting
our calculations of the sensitivity and resolution required throughout
the chapter. We draw attention to other chapters in this volume that
cover aspects of radio AGN science, in particular the chapters by
\cite{Baczko01.2026.SKA}, \cite{Baldi01.2026.SKA}, \cite{Pandey-Pommier01.2026.SKA, Pandey-Pommier03.2026.SKA}, and \cite{Panessa01.2026.SKA}
which include discussion of radio AGN VLBI with the SKA.

Throughout the chapter we refer to the power-law radio spectral index $\alpha$,
defined in the sense that $S_\nu \propto \nu^{-\alpha}$. For a power-law
electron energy distribution, $N(E) \propto E^{-p}$, the spectral
index $\alpha$ is given by $\alpha = (p-1)/2$.

\section{Observations of radio galaxies with the SKA}

\begin{figure}[t]
  \includegraphics[width=\linewidth]{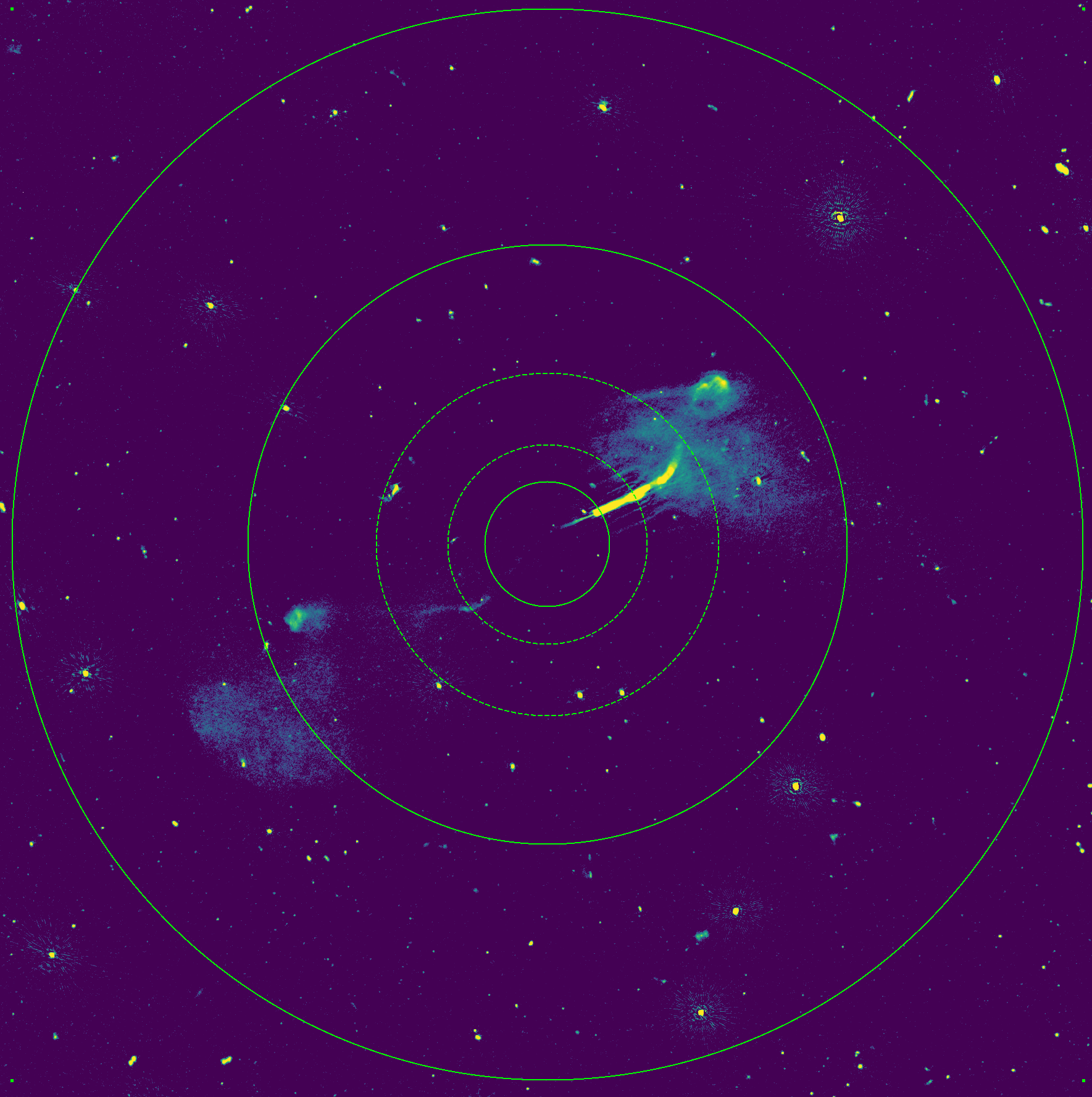}
  \caption{LOFAR (LoTSS: \citealt{Shimwell+26}) view of the nearby degree-scale radio galaxy
    NGC 6251 at 144 MHz with 6-arcsec resolution. Many other smaller
    radio galaxies can be seen in the same field. The image shows a
    $1.5^\circ \times 1.5^\circ$ field of view (FOV) matched to the
    FOV of SKA-Low at 300 MHz, the frequency at which Low will achieve
    the same resolution as LOFAR. Circles show, from largest to
    smallest, the FWHM FOV of Low at 300 MHz, which is similar to that
    of Mid band 1 at its central frequency, and the FOV of Mid bands
    2, 3 and 4 (dotted lines as bands 3 and 4 will not be available
    initially), and 5a at their central frequencies. Observations of
    the largest sources will be increasingly restricted by the FOV and
    the shortest available baseline as they move to higher frequency
    and higher resolution.}
  \label{fig:6251}
\end{figure}

As discussed above, radio galaxies are synchrotron emitters, which
means that their radio emission is broad-band, extending in many cases
from the lowest to highest frequencies available. Thus the entire
frequency range from 50 MHz to 15 GHz offered by the SKA can be used
for this type of work. However, resolution is a key limiting factor.
As we will outline below, much of the insight into radio AGN physics
comes from resolved studies of the population, and observations show
\citep{Sweijen+25} that the angular sizes of about half the studied
population of radio galaxies in one LOFAR deep field with flux density
$>0.6$ mJy are less than a few arcsec. The resolution of SKA-Low
(hereafter Low) will range from around 20 to around 3 arcsec across
the band (from lowest to highest frequency) while that of SKA-Mid
(hereafter Mid) will range from around 1.5 arcsec to 0.03 arcsec
across bands 1-5. Thus it will only be possible to resolve the bulk of
the population at the highest frequencies using Mid, while resolved
studies with Low will tend to cover a luminous, physically large
subset of targets. Wide-area surveys with Low (at the
high-frequency end to avoid the confusion limit) should however allow
the selection of a large population of AGN, with a sky density around
1000 deg$^{-2}$ \citep{Hardcastle+25}. Because of its relatively small
(few arcmin) field of view, Mid (particularly at band 5) will likely
be most useful for targeted observations of individual objects or for
deep surveys of smaller sky areas. The big gains in this area will
come from the SKA's sensitivity and image fidelity compared to other
facilities, which will enable us to build up observations of
statistically meaningfully samples much more quickly than hitherto.

Physical size is also a factor. Powerful classical radio galaxies are
often physically large, extending to scales of hundreds of kpc to Mpc
scales \citep{Laing+83,Bridle+Perley84}, and these sizes correspond to
large angular sizes (Fig.\ \ref{fig:6251}). To image their extended
structure and measure physical size and other parameters of their
lobes, such as their Fanaroff-Riley classification
\citep{Fanaroff+Riley74} arcsec-scale resolution is often sufficient,
including at very low radio ($\sim$58 MHz) frequencies
\citep{Boxelaar+25}, while the imaging of these largest-scale
structures at high frequencies will be limited by the Mid primary
beam and by the SKA's shortest baselines (which, however, are very
short, of the order of twice the diameter of the individual antennas
for Mid, meaning that structures of order half the primary beam in
size can be imaged).

Radio galaxies all launch bipolar jets, but as is well known, their
morphology is related to their luminosity: the most powerful objects
tend to have a Fanaroff-Riley type II (FRII) morphology
\citep{Fanaroff+Riley74,Bernie2021} showing faint or invisible emission related
to the jets but bright hotspots at the end of well defined lobes,
whereas in the classical Fanaroff-Riley class I (FRI) morphology
bright inner jets feed into lobes or plumes. This morphological
classification and its relationship to luminosity was based on
low-sensitivity, low-fidelity observations of small samples of bright
objects, and precursor work has shown us that we can expect objects
that do not fit this pattern \citep{Capetti+17,Mingo+19} as well as much more
complication in the low-surface-brightness structures that will be
revealed with the SKA. These classifications (or the structures
responsible for them) most likely also evolve with redshift, something
that can only be properly explored with the SKA.

Below we list some observational projects that
will be enabled by the SKA's capabilities.

\subsection{Distribution of source sizes and luminosities}

As noted above, observations of the distributions of radio galaxies on
the power-linear size ($PD$) plane are fundamental to the study of the
life cycle of radio galaxies. Together with the evolving radio
luminosity function, they provide the large-scale observational
constraints on any model that links the radio AGN population to the
accretion history of supermassive black holes and the evolution of
their host galaxies. The SKA's potential contribution in this area was
already discussed by \cite{Kapinska.etal2014} and the reader is
referred to that paper for further details. Here we just note that the
change in design of the SKA relative to what was under consideration
in 2014 does affect its capabilities in that area. Wide-area surveys
with Low, as discussed above, will give an unrivalled deep view of the
overall population, particularly when combined with EMU data at higher
frequencies and with the deep optical data that will be available over
the Southern hemisphere in the coming decades, and will allow the
selection of sources for followup with Mid along the lines discussed
further below. We can expect Low to see much the same range of luminosity of AGN as LOFAR currently does, i.e. between around $10^{21}$ and $10^{30}$ W Hz$^{-1}$ at 150 MHz \citep{Hardcastle+25} But we can no longer expect the wide-area surveys to
give us sub-arcsec resolution over large areas, and so to resolve the
bulk of the radio AGN population; objects smaller than 10-100 kpc (for a plausible distribution of redshifts) will remain unresolved with Low, and this constitutes a significant fraction, up to 80\%, of an AGN sample \citep{Hardcastle+25,Sweijen+25}. Size and morphological information will need to be
provided in the SKA1 era by a combination of observations with Mid
over smaller fields and other instruments, such as LOFAR, or by VLBI. It is
certainly possible to imagine a high-frequency survey with Mid that
would give high resolution over intermediate areas --- comparable, for
example, to the
tens of square degrees currently provided by the MeerKAT International
GHz Tiered Extragalactic Exploration survey (MIGHTEE: \citealt{Hale+25})
but conducted at band 5a, where the resolution will be around 0.1 arcsec.
Covering the 20$^\circ$ MIGHTEE sky area to the same depth as
MIGHTEE but at band 5a, taking account of a typical synchrotron
spectral index, would require a few hundred hours of Mid observing
time, assuming that a sensitivity of 0.7 $\mu$Jy beam$^{-1}$ would be achieved over around 0.15
deg$^2$ in one hour. Such a survey would resolve the vast majority of radio AGN in
the field \citep{Sweijen+25} and would also greatly improve the ability to separate AGN
emission from that due to star formation \citep[cf.][]{Morabito+22}.

\subsection{Spectral ageing studies}
\label{sec:specage}

As a population of electrons in a radio galaxy radiates, higher-energy
electrons radiate faster than lower-energy ones, which leads to a
characteristic steepening of the radio spectrum
\citep[e.g.][]{Jaffe+Perola73}. It has long been known that the
synchrotron spectra of FRII radio galaxies systematically steepen
with distance from the hot spots \citep{Alexander+Leahy87}, while
FRI radio galaxies show spectra that steepen with distance along the
jets \citep{Burch77} and then, if lobes are present, with distance
into the lobes as for FRIIs. The steepening is interpreted as evidence for
`synchrotron ageing', whereby the synchrotron energy loss process
depletes high-energy electrons in the source as they move away from
the acceleration site \citep{Pacholczyk70,Jaffe+Perola73,Leahy91}. If
we are really observing the effects of synchrotron ageing, and if the
magnetic field strength is known, or can be estimated, e.g. from
inverse-Compton observations \citep[e.g][]{Ineson+17} then in principle
resolved images of the radio spectra allow us to measure plasma age as
a function of position in the source, which allows the determination
of quantities like total source age and expansion speed, and also
allows a timeline to be established for the source's phase(s) of
activity. Different models of radiative loss make different
predictions which can in principle be tested with a high-enough
quality radio spectrum.

While broad-band radio observations have established the reality of
spectral ageing at low frequencies
\citep{Lal+Rao04,Lal+Rao07,Sakelliou+08,Harwood+13, Jamrozy.etal2008}, high-frequency
observations are important in this context because the spectral age
probed by a frequency $\nu$ goes as $\nu^{-1/2}$. Thus, in principle,
observations at the high end of frequency probe electrons an order of
magnitude younger than those seen in the low end of frequency.
Relative to existing studies with the VLA, GMRT or LOFAR, SKA spectral
studies will be able to produce well-sampled, matched-resolution
images across two decades in frequency for appropriately sized sources,
due to the SKA's unrivalled $(u,v)$ coverage. For compact, young
sources, resolved spectral ageing studies are likely {\it only} to be possible
with the high frequencies of Mid \citep[e.g.][]{Heesen+14}, although
these will need to be tied to low frequencies to estimate injection
indices, and care will be needed to take account of low-frequency
turnovers that are known to appear in higher frequencies in smaller
sources \citep{ODea+Baum97} and which will require modelling in terms of
free-free \citep[e.g.][]{Bicknell+18,Young+25} and/or synchrotron self-absorption.

Broad-band spectral studies have the capability to answer three key questions:
\begin{enumerate}
  \item What are the spectra of aged regions? Are they consistent across a broad spectral range with theoretical models?
  \item Are the spectral ages derived from detailed imaging consistent
    with dynamical ages derived from models of radio source expansion
    and observations of radio source environments?
  \item Are the particles responsible for the low-frequency emission entirely co-spatial with those responsible for the high-frequency emission as seen in FRII radio galaxies?
\end{enumerate}

The first of these questions is important because it gives insight
into the micro-physics of the radio structures --- in spite of many
studies investigating this question, GHz-frequency work does not provide a definitive answer to the question of whether pitch-angle scattering is effective during loss as expected in the models of \cite{Jaffe+Perola73}, for example \citep[e.g.][]{Harwood+13}. 
High-resolution, high-fidelity, truly broad-band images should answer this by unambiguously locating the cutoff in the synchrotron spectrum at high energies if it exists, and maybe allow us to probe inhomogeneities in the magnetic field strengths as well \citep{Tribble93,Hardcastle13}.

Both the second and third questions are fundamental to the interpretation of spectral age measurements. The so-called `spectral age problem' \citep{Eilek96} is the fact that spectral ages seem to be significantly younger than the expected dynamical ages where these can be modelled. 
Although the details of this discrepancy depend on factors like the unknown lobe magnetic field strength (and its history), the problem persists even when the best possible field strengths are used \citep{Harwood+13,Mahatma+20}. Possible solutions include mixing of plasma of different ages in the lobes \citep{Turner+18,Jerrim+25}, and/or {\it in situ} particle acceleration in the lobe, which should be identifiable as a spectral {\it flattening} in certain parts of the lobes at high frequency \citep{Lal+Rao07}. 
Existing low-resolution, low-sensitivity studies of small samples of lobes at high radio frequencies cannot test this model \citep{Hardcastle+Looney01} but SKA broad-band observations will enable us to establish the spectral properties of morphologically defined subsamples of RLAGN on a {\it statistical} rather than an individual basis. 
In addition to solving the spectral age problem and helping us to calibrate observed spectral ages as a constraint on radio source models, direct evidence for {\it in situ} leptonic particle acceleration in the lobes would give a previously unavailable insight into lobe micro-physics and might connect to models of the ultra-high-energy cosmic ray population \citep{Hardcastle10,Matthews+19}.

With high sensitivity and image fidelity, it is now routine to fit
spectral models to each individual pixel of an image of a source
\citep{Harwood+13}, and the broader the frequency range available, the
better the age constraints are. The key advance will be to build up
systematic, consistent spectral age information for large samples. For
example, \cite{Pinjarkar+23} used LOFAR, superMIGHTEE \citep[upgraded
  GMRT;][]{Lal2025} and MIGHTEE (MeerKAT) data at 144-1400 MHz to
determine spectral ages for a sample of resolved sources in 3.5
deg$^2$ of the XMM-LSS deep field; but they were limited by their data
to 10 arcsec resolution and to sources brighter than tens of mJy at
144 MHz. At the lowest-frequency end of Mid, where the resolution will
be about 1.5 arcsec, a comparable deep-field survey could reach higher
sensitivity within an hour. In band 5, one hour per pointing (covering
several sources) would enable imaging of objects up to ten times
fainter and six times smaller than those studied by
\cite{Pinjarkar+23}, providing matched-resolution coverage from 350
MHz to 15 GHz for a sample tens to hundreds of times larger than the
current state of the art.

\subsection{Magnetic fields and rotation measure in lobes}
\label{sec:polarization}

The radio emission from radio galaxies is always intrinsically
polarized, with an intrinsic direction which gives information on the
mean magnetic field direction along the line of sight, and a
fractional polarization which reflects the degree to which magnetic
fields are aligned versus tangled on scales much smaller than the
beam. However, this is complicated by Faraday rotation due to thermal
electrons, which can be in the halo of the Milky Way, in the local
environment of the radio galaxy but external to it, or even internal
to the lobes. Unresolved Faraday rotation structure in front of the
lobes can give rise to depolarization \citep{Burn66}, which is
frequency-dependent since the angle of Faraday rotation goes as
$\lambda^2$, but if the Faraday screen can be resolved then only
rotation will be seen. Faraday rotation internal to the lobes due to
internal thermal electrons will lead to depolarization, independent of
resolution, at low
frequencies ($\nu < c\sqrt{Cn_\mathrm{th}|B_\parallel| D}$, where
$n_\mathrm{th}$ is the number density of thermal electrons,
$B_\parallel$ is the magnetic field strength along the line of sight,
$D$ is the depth through the source and $C = 2.62 \times
10^{-13}$ T$^{-1}$).

If the interest is in studying the magnetic field structure internal
to the lobes, then working at high frequencies (band 5) will offer
both the highest resolution and the lowest Faraday effects. Past
studies of individual objects or small samples at high frequency with
instruments such as the un-upgraded VLA required many tens of hours to
build up the uv coverage required through multiple configurations. A
classic study of this type by \cite{Bridle+94}
reached noise levels of 20 $\mu$Jy beam$^{-1}$ at 5 GHz and imaged
polarization in a modest sample of quasars with 0.35-arcsec
resolution. A ten-minute observation in band 5a will achieve the same
{\it surface brightness} sensitivity with a beam 7 times smaller in
area and with superior image fidelity. This would allow the generation
of samples large enough for robust comparisons with MHD simulations,
as well as probing the small-scale field structure in jets and
hotspots (see below).

However, there is also substantial information in the Faraday rotation
and/or depolarization itself, and studying that requires us to go to
lower frequencies. Faraday rotation from an external medium gives us
information about the product of the thermal electron density and the
magnetic field vector component along the line of sight: this means
that even if the thermal electron density is reasonably uniform, the
rotation measure can vary on small scales from point to point due to
the turbulent field in the intergalactic medium (where the definition
of `small scales' depends on the poorly known MHD turbulence scales in
the external medium). At frequencies and resolutions that are too low,
sources can thus be completely depolarized by beam depolarization
effects \citep{OSullivan+23}, but in between (at frequencies between
hundreds of MHz and a few GHz for typical radio galaxy environmental
conditions) the Faraday screen can be partly or wholly resolved
\citep[e.g.][]{Guidetti+11,Baidoo+23} and there is a sweet spot where
information can be extracted about the large-scale environment without
the need for expensive X-ray observations \citep{Jerrim+24,Stimpson+25}. Mid in
bands 1 and 2 will be good in particular for the large-scale
structures of radio AGN in poor environments, where there are few
useful probes of the thermal electron density. Broad-band polarimetry
at these frequency ranges is hardly explored, but we can compare with
the pre-upgrade VLA imaging used by \cite{Guidetti+11}, where
observations of duration of several hours with resolution of 1.5
arcsec and sensitivity of 25 $\mu$Jy beam$^{-1}$ over a 50-MHz channel
were used. In SKA band 2, the same sensitivity but with twice the
resolution would be achieved in a 5-minute observation, generating 16
images of the same channel width to give an in-band rotation measure
constraint. Much finer channel spacing could of course be used for
rotation measure synthesis. Observations at the lowest frequencies
possible (Mid band 1 or Low) will give the best constraints on any
internal thermal electron content, and should be carried out on large
scales where the external Faraday rotation is low, though it will
likely be challenging to distinguish from external depolarization
\citep{Tribble91,Stimpson+25}.

\subsection{Particle acceleration locations in FRIs and FRIIs}

Jets of FRI and FRII sources interact differently with their
environments \citep{LalHardcastle2008}. It is generally believed that all jets start off
highly relativistic, and hence supersonic with respect to their
internal sound speed, but that FRI jets decelerate by
entrainment of thermal material, either from stellar winds or from the
external environment, which injects mass into the jet and so
decelerates it by conservation of momentum
\citep{Bicknell84,Laing+Bridle02,LalKraft2010,LalKraft2013,Perucho+14}. Dissipative processes
once the jet becomes trans-sonic cause particle acceleration and
brightening in the inner parts of the jet, typically 1-10 kpc for
powerful sources such as 3C\,31 \citep{Laing+Bridle02}. By contrast, FRII jets are
powerful enough to escape from their galactic environment without
significant deceleration, and so remain supersonic until the end of
the jet where they terminate in a jet termination shock. The bright
radio emission at the ends of FRII sources, which originates in the
forward and/or reverse shock associated with jet termination, is
thought to be the site of particle acceleration in FRIIs.

In pre-SKA instruments, the particle acceleration locations in both
types of object, which have structure on small scales, have been hard
to study with high resolution and sensitivity, as instruments like the
VLA do not provide in a single observation the long and short
baselines needed to image hotspots and inner jets in the presence of
bright diffuse emission from lobes and plumes. The SKA will be able to
image these structures in bulk at high resolution and with
polarization capabilities, probing models of particle acceleration
physics. High-resolution images of these regions at multiple
frequencies will probe the `injection index', the spectral index of
material injected at the hotspot, which is a crucial ingredient in
models of spectral ageing, and test whether it is consistent with the
expectation from shock models. Hotspots in particular appear to be
transient features, which evolve on a short timescale compared to the
lifetime of a source, and the detailed structure and spectral ageing
history of multiple hotspots can shed light on models of jet
precession \citep{Horton+23}. The morphology of jets and hotspots may
also differ between low and high-power FRIIs. Jets propagating through thermal ambient
medium (for normal radio galaxies) and non-thermal ambient medium (for
the inner jets of episodic radio galaxies) may have subtle difference
in injection indices, hotspot speeds and efficiency of particle
acceleration that can only be studied with the high resolution and
high sensitivity observations that will be available with the SKA. As
with other cases, the SKA will be able to build up large samples of
high-resolution images of such objects with very high speed compared
to previous work. Hotspots of FRIIs, because these objects are
generally distant, need sub-arcsec resolution to study sub-kpc
structure, and so are good targets for observations at band 5. For
example, \cite{Hardcastle+98} used observations of bright 3C sources
at a typical frequency of 8 GHz and a resolution of 0.25 arcsec to
measure properties of hotspots at a single frequency from maps with
rms noise levels of tens of $\mu$Jy beam$^{-1}$, accumulating a sample
of around 50 objects, based on observations spanning many hours of
observation and many years of time due to the VLA's configuration
cycle. Using 5-minute snapshots in SKA band 5a, a sample ten times
larger, with a sensitivity ten times higher and a resolution twice as
good (plus in-band spectral index information), could be obtained in a
two-day observing campaign after selection from a wide-area survey.

\subsection{Jet speed and structure}

The speed of relativistic jets can be measured by observations of
Doppler boosting and, for polarimetric observations, aberration, if we
assume that the jet and counterjet in a particular system are
intrinsically identical. Observations of jet deceleration and velocity
structure in FRI sources are a key piece of evidence for the standard
model of the FRI/FRII difference, as mentioned above. The sensitive,
high-resolution polarimetric observations needed to do this for small
samples of FRIs with the VLA \citep{Laing+Bridle14} could be generalized to
much larger samples with the SKA to enable the study of the dependence
of the jet kinematics on the sources' environment, host galaxy
properties and jet power, and hence test the entrainment model for jet
kinematics in detail.

Understanding the kinematics of FRII jets is much harder because the
jets are typically much fainter, unless strongly beamed, and
counterjets (the Doppler-suppressed jet receding from us) are often
not detected in available imaging. Constraints come from the
distribution of jet sidedness \citep[e.g.][]{Wardle+Aaron97} and jet
prominence
\citep[e.g.][]{LalHardcastle2008,Mullin+Hardcastle09,LalKraft2010},
which give beaming speed estimates corresponding to $0.5 < \beta <
0.8$, consistent with the idea that these jets remain relativistic.
However, these beaming speeds for jets (in terms of Lorentz factor,
$\Gamma \sim 1.5$) are much lower than the corresponding speeds for
cores ($\Gamma \sim 10$) or to the values directly measured from VLBI
observations of bulk speeds in quasars ($\Gamma > 20$); yet there is
no evidence of the energy dissipation that would be observed if the
jets decelerated from bulk $\Gamma \sim 10$ to $\Gamma \sim 1.5$ on
kpc scales. The generally accepted explanation for this is that the FRII jets
have velocity structure, and that the material seen on large scales is
a slow-moving `sheath' around a fast jet core which may carry most of
the energy\footnote{\cite{Konar+Hardcastle13} argue independently that
the observed physics of double-double radio galaxies requires that a
highly relativistic jet persist out to the hotspots, consistent with
the spine-sheath model.}. As noted above, velocity structure is
already seen in Doppler boosting analysis of FRI jets, but if this is
the case, we might expect to see resolved structure in total intensity
and polarization in the jets of FRIIs as well when observed at the
highest available resolution, e.g. with SKA band 5b. In addition, such
observations should have the image fidelity to image counterjets in
many more objects than at present (the limiting factor is likely the
image fidelity needed to pick out the counterjets in the presence of
complex filamentary structure in the lobes, rather than raw
sensitivity) which gives additional constraints on beaming.
Observations of 3C\,353, a nearby low-power FRII source
\citep{Swain+98} are consistent with such a model but only the SKA can
give the combination of high ($\sim 100$-pc) resolution and
sensitivity to build up an unbiased sample for this kind of study.

\section{Physics of Radio Galaxies: Episodic Radio Galaxies}
\label{sec:erg}
\begin{figure}[t]
  \begin{center}\includegraphics[width=0.75\linewidth]{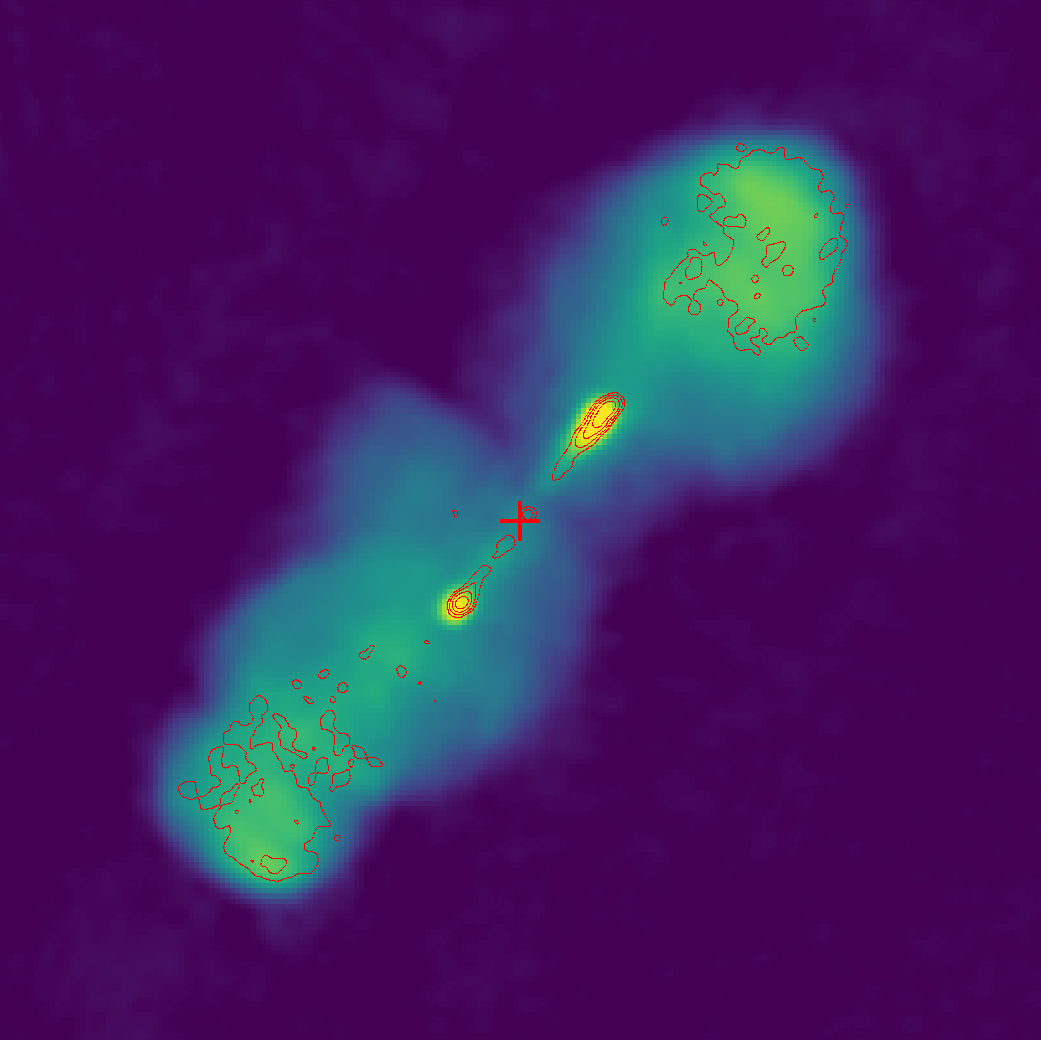}
    \end{center}
\caption{A double-double radio galaxy, J1158+2621 (4C\ 26.35), is 
shown in this image, which has a $5 \times 5$ arcmin field of view. The plus sign in red indicates the central position
(11:58:20.1400, +26:21:12.00) of the optical host galaxy. The colour scale
is the image from the LoTSS survey image data at 6 arcsec resolution while the contours 
are from the 1.4-GHz GMRT imaging of \cite{Konar.etal2013}, who carried out a
detailed spectral ageing analysis. The contour levels are
1,2,4,8 mJy beam$^{-1}$ for a resolution of 2.6 arcsec, while the logarithmic
colour scale shows a lowest surface brightness of 1 mJy beam$^{-1}$
with 6-arcsec beam at 144 MHz.}   
  \label{fig:DDRG_J1158}
\end{figure}

Episodic jet forming activity of AGN is quite ubiquitous in the
Universe \citep{Schoenmakers+00, Saripalli.etal2003, Saikia.etal2006,
  Safouris.etal2008, Konar.etal2013, Konar+Hardcastle13,
  Nandi.etal2019, Saikia&jamrozy2009, Mahatma.etal2023, Morganti2024,
  Dabhade.etal2025}. If radio loud AGN are observed with two episodes,
i.e., two pairs of lobes with a common core, they are called
double-double radio galaxies. Similarly, objects with three pairs of
lobes with a common core are called triple-double radio galaxies.
Objects which host a very small inner double, often seen as a
steep-spectrum core, and so are presumably an
early phase of the evolution of a double-double radio galaxy, fall
under the general label of episodic sources.
Since they presumably trace the same underlying physics of an
interruption in the jet energy supply (Section \ref{sec:intro}) we will
use the terms episodic, double-double, and triple-double synonymously
throughout this chapter. For the double-double radio galaxies, the
younger lobes inflated by the re-started jets in the current episode
and lying between the two older lobes of the previous episode are
called the inner double. The outer older lobes are called the outer
double. Fig~\ref{fig:DDRG_J1158} shows an example in which an inner
double is embedded in the diffuse relic plasma of outer lobes created
in the previous episode. The episodic behaviour does not have any
known dependence on the host galaxy of the AGN
\citep[e.g.][]{Mahatma.etal2023}. It is yet to be understood whether
the host galaxy of the AGN has any role to play in deciding the
timescale of the active and quiescent phases.

%Studies of jet and lobe dynamics with the help of broad band radio
%spectra and spectral ageing analysis have been largely done for what
%are called double-double radio galaxies. As an example, a
%double-double radio galaxy is shown in Fig~\ref{fig:DDRG_J1158}. The
%grey and contour images are from two different surveys (see the
%caption of the figure). The image in Fig~\ref{fig:DDRG_J1158} also
%illustrate how two different surveys of completely different
%resolutions help us in identifying episodic radio galaxies by
%detecting outer and inner structure. Objects with very small-scale
%inner double sources, which are now known as `radio galaxies with
%steep spectrum cores' may be found to be classical double-double radio
%galaxies with the high resolution and sensitivity of the SKA.

Episodic behaviour of the radio loud AGN is responsible 
for the production of remnant radio galaxies and restarted jets. A lot 
of information related to duty cycle and jet dynamics comes from the 
remnant lobes (or, relic lobes) and the restarted jets through spectral 
ageing analysis. In this section, we discuss the SKA's potential to improve studies of duty
cycles and jet dynamics based on episodic AGN. We defer our detailed discussion on remnant 
radio galaxies to Section~\ref{sec:remnant}.

%%%%%%%%%%%%%%%%%%   DDRG Image  %%%%%%%%%%%%%%%%%%%%%%%%%%%

%%%%%%%%%%%%%%%%%%%%%%%%%%%%%%%%%%%%%%%%%%%%%%%%%%%%%%%%%%%%

%%%%%%%%%%%%%%%%%%%%%%%%%%%%%%%%%%%%%%%%%%%%%%%%%%%%%%%%%%%%%%%%%%%%%
\subsection{What causes and controls jet activity in radio loud AGN?}

As noted above and in Section \ref{sec:intro}, it is now clear that jet activity
in radio loud AGN is episodic. The primary evidence for this comes
from observations of the radio lobes created by a previous episode of
jet activity \citep{Schoenmakers+00,LalHardcastle2008}. However, it is
not clear a) what exactly causes the episodic behaviour, b) whether
all radio loud AGN will exhibit the episodic activity, c) whether
currently radio quiet AGN will some day start jet activity, d)
whether the episodic jet activity is connected with a change of
accretion state (low-hard and high soft states) as in stellar mass
black hole binaries, and e) whether the feedback loop between
supermassive black hole and thermal ambient medium can be causally
connected to the episodic behaviour in jet activity. Addressing some
of these questions needs X-ray, optical, mm (e.g., CO line) 
and cm (e.g. HI line) observations; however,
radio observations with high resolution will definitely provide key
information. For example, one of the crucial constraints for inferring
the cause of episodic radio galaxies is the range of observed duty
cycle, as any model has to reproduce this in order to be considered
successful. The duty cycle of jet activity, or jet duty cycle, is
given by
\begin{equation}
\text{Jet~duty~cycle} = t_{\text{active}}/(t_{\text{active}} + t_{\text{quies}}),
\label{dutycycle_eqn} 
\end{equation}
where $t_{\text{active}}$ and $t_{\text{quies}}$ 
are the timescales of the active and quiescent phase 
of jet activity respectively. In the next section we consider how this
can be constrained by radio observations.

\subsection{What are the values of  jet duty cycles for AGN?}

Constraining jet duty cycle has been attempted only for a handful of
sources by \cite{Konar.etal2013} through spectral ageing analysis for
FRII sources. They did not quote the values of the jet duty cycle
as per the definition given above, but rather estimated the
timescales of the active phase ($t_{\text{active}}$) and quiescent phase
($t_{\text{quies}}$) of jet activity. The jet duty cycle can thus in
principle be estimated from their results for a handful of sources\footnote{These tend to be FRII sources because it is easier to identify the
characteristic double-double morphology, although FRI sources can also
be shown to be episodic in a few well-studied objects, such as the
multiple outbursts seen in the nearest radio galaxy, Centaurus A
\citep{Morganti+99}.}. In addition, \cite{Konar.etal2013} obtained only 
the upper and lower bounds on $t_{\text{innd}}$. The higher the resolution 
and sensitivity, the better we can constrain $t_{\text{innd}}$ and hence 
the duty cycle, as we discuss below.

We can make use of spectral ageing analysis (Section \ref{sec:specage}) to 
constrain the age of the outer double ($t_{\text{outd}}$) 
and the age of the inner double ($t_{\text{innd}}$) 
and the time period of the quiescent phase ($t_{\text{quies}}$)
of jet activity between two episodes.
The formulae for timescale of active phase and quiescent 
phase derived by \cite{Konar.etal2013} are 
\begin{equation}  
t_{\text{active}} = t_{\text{outd}} - (t_{\text{jet}} + t_{\text{ws}})        
\end{equation} 
and        
\begin{equation}
t_{\text{quies}} = (t_{\text{ws}} + t_{\text{jet}}) - t_{\text{innd}}               
\end{equation}  
respectively; where $t_{\text{active}}$ and $t_{\text{quies}}$ are the
timescale of active and quiescent phase of the jet activity of the
radio loud AGN respectively, $t_{\text{outd}}$ and $t_{\text{innd}}$
are the spectral ages of the inner and the outer doubles respectively,
$t_{\text{jet}}$ is the jet travel time of the radio emitting plasma
for the outer double, and $t_{\text{ws}}$ is the spectral age of the
`warm spot', i.e., the time elapsed between the moment when the last
jet material was dumped at the outer hotspot (now the `warm spot' at
the end of the outer lobe) and the moment of observation.

Though \cite{Konar.etal2013} had good estimates of $t_{\text{outd}}$, 
$t_{\text{jet}}$ and $t_{\text{ws}}$, they had only limits on
$t_{\text{innd}}$. This is because the integrated spectra of 
the inner doubles were power-laws within the observable 
range (few hundred MHz to around 8 or 15 GHz) in their observations to
the accuracy permitted by their data. 
No curvature or breaks were observed in the 
integrated spectra of the inner doubles.  Moreover, because of 
the limited resolution and sensitivity of the GMRT and JVLA, 
they could not estimate the ages at various parts of the lobes 
of the inner doubles of their samples, although this was possible for
the outer lobes because of their larger size. 
Integrated spectra cannot give us accurate ages even if 
the curvature is detected, as the non-thermal plasma injected by 
the jets into the lobes is injected at different times and with
different values of the magnetic field strength (see also the
discussion of the spectral age/dynamical age problem in Section \ref{sec:specage}).

The study of the inner lobes of episodic radio galaxies (e.g.,
double-double radio galaxies) will be much easier with the SKA. If the
inner lobes can be imaged with sub-arcsecond resolution at several
frequency bands from the lowest to the highest possible, the spectral
ageing analysis will be greatly improved. The wide frequency coverage
of the SKA along with its high sensitivity and resolution will greatly
improve spectral ageing analyis, as discussed above (Section
\ref{sec:specage}). For the inner doubles, although
\cite{Konar.etal2006}, \cite{Konar.etal2012}, \cite{Konar.etal2013}
and \cite{Jamrozy.etal2007} did not find any curvature in the
integrated spectra of inner double of their sample sources, it is
likely that curvature will be found for the spectra of the inner-lobe
strips which are near the cores, which will help to constrain the ages
of the inner doubles instead of giving limits on their ages. For some
selected double-double radio galaxies like J1453+3308
\citep{Konar.etal2006}, 4C29.30 \citep{Jamrozy.etal2007}, J0116-4722
and J1158+2621 \citep{Konar.etal2013}, the inner doubles will be well
resolved at 1 arcsec or sub-arcsec resolution which will be available
at frequencies higher than 500 MHz. That means that in all the Mid
array bands the inner doubles of those sources will be well resolved
and spectral ageing analysis in various strips of the inner lobes will
be possible using the Mid array. Sensitivity will not be an issue at
all. In J1453+3308 \citep{Konar.etal2006}, we can see that in a 5-GHz
image the backflow part of the lobes was detected down to a
sensitivity level of 50 $\mu$Jy beam$^{-1}$. As Mid in band 5 will
reach $\mu$Jy beam$^{-1}$ noise levels in a ten-minute observation,
all of the diffuse parts of the inner lobes are expected to be
detected in a plausible SKA observing campaign. At frequencies lower
than 5 GHz, the diffuse emission will be brighter, and hence is
expected to be easily detected. At frequencies higher than 5 GHz, we
cannot predict very firmly how much diffuse emission of the lobes will
be detected, as the diffuse emission of the inner lobes is expected to
suffer synchrotron losses. But irrespective of whether we can predict
the diffuse emission at frequencies higher than 5 GHz, we can firmly
state that spectral ageing analysis for various strips of the inner
lobes, or even per pixel, will be possible with the help of the Mid
array. Mid band 1 and Low will also be useful for spectral ageing
analysis for the outer doubles and constraining the injection spectral
indices of inner and outer doubles, for which we can use the
integrated flux densities. Usually, we assume that the injection
spectral index remains similar for the entire active phase of the jet
activity. We can even afford to be skeptical about this assumption in
the SKA era, however, as we can constrain the spectra of various
strips of inner and outer lobes with higher sensitivity and/or carry
out a pixel by pixel analysis. Thus, we can constrain the injection
spectral index for many regions of the lobe separately and check if
that is similar throughout the active phase of jet activity of radio
galaxies. Since there will be good $uv$ coverage for small baselines,
the detection of diffuse emission for either the inner or outer
doubles of double-double radio galaxies will not be an issue. There
are already examples of variation of spectral injection spectral
indices from one phase to the other in work by \cite{Brienza.etal2020,
  Brienza+21} and \cite{Rajpurohit.etal2024}. This may be due to
variation of jet power over the lifetime of a single episode.

\subsection{Hidden signatures of episodic activity in radio galaxies: relic lobes and steep spectrum cores}
\label{subsec:3.3}

As noted above, radio galaxies are episodic and for FRII radio
galaxies we can observe their episodes as multiple lobes, giving rise
to a description of them as single double, double-double, triple
double etc. The FRII radio galaxies which are observed as `single
doubles', i.e, only a single phase of activity is visible, are
expected to cease their jet activity at some point in the future and
be in a quiescent phase for some time before the restart their jet
activity (see further below, Section \ref{sec:remnant}). It is also
plausible that something that now appears as a simple active radio
galaxy might have finished its previous episode of jet activity so
long ago that we do not observe the remnant lobes from a previous
outburst as their surface brightness has dropped below the sensitivity
level of the current radio telescopes. The work of searching for these
remnant radio galaxies has already been started with the SKA precursor
telescopes (see \cite{Brienza+17, Mahatma+18, Jurlin+20, Shabala+20}
and many more) and a large number of candidate remnants have been
discovered. With the SKA, with its extremely high surface brightness
sensitivity, such large-scale diffuse emission around radio galaxies
is expected to be discovered easily. As \cite{Kapinska.etal2014}
pointed out, there is a `hidden world of secret lives of radio
galaxies' that we can discover with instruments sensitive to low
surface brightness, low-frequency emission. A preview of this is given
by LOFAR, where we see several well-studied objects that show
signatures of steep-spectrum, low-surface brightness emission
interacting with the external medium
\citep[e.g.][]{Hardcastle+19b,Brienza+21} but Low should give us many
more, as Low will have a factor of 10 better sensitivity than LoTSS
(and reach its confusion limit) with only 1 hour of integration time.

Another class of radio galaxies are those that show an unresolved 
but steep-spectrum or GPS core
with the ordinary resolution of existing telescopes (in constrast to
the flat-spectrum cores normally seen, which are the self-absorbed
bases of the inner jets). With higher
resolution sometimes these cores are resolved into mini-doubles,
providing evidence for an additional epoch of jet activity that has
only recently begun. An
illustrative example of such a source is B0925+420, published by
\cite{Brocksopp.etal2007}. They imaged this source with the 4.9 GHz
and 8.4 GHz data both taken in VLA B-array. The innermost-double is
clearly resolved with the VLA in both the 4.8 GHz and 8.4 GHz images,
i.e. at arcsec resolution. Therefore,
it will be worth investigating the steep spectrum cores of radio
galaxies with the SKA.
\cite{Godambe.etal2009} investigated three relic sources, namely
J0139+3957, J0200+4049 and J0807+7400, which have steep spectrum cores.
The core spectrum of J0139+3957 is steeper at frequencies above 3 GHz,
and gradually flattens to zero spectral index at lower frequencies.
The core spectrum of J0200+4049 is steeper at frequencies above 1 GHz,
and gradually flattens to zero and becomes slightly inverted at lower
frequencies. These objects are therefore reminiscent of 
Giga-Hertz Peaked Spectrum (GPS) sources but exist inside large-scale 
lobes. The core of the other source J0807+7400 has also a steep
spectral index between 300 MHz and 1.0 GHz. Beyond that it has a
complicated spectrum: at lower frequencies than 300 MHz, there seems to
be some absorption, while at frequencies higher than 1 GHz, the spectrum seems
to be affected by variability. The highest resolution of their images
was a few arcseconds and the cores were not resolved. 
Similar work in a larger sample, where prominent and steep spectrum 
cores have been interpreted as restarted activity, has been carried
out by \cite{Nair.etal2024}. Many more such
sources may be discovered with mini
doubles inside the steep spectrum cores using the SKA, which will have
the ability to measure in-band spectral index at the same time as
carrying out high-resolution imaging in a single band. \cite{Brocksopp.etal2007}
detected the diffuse lobe of the innermost double in the radio source
B0925+420 with a sensitivity level of 22 $\mu$Jy beam$^{-1}$ in
arcsec-resolution VLA observations. With an order of magnitude more
sensitivity, as provided by Mid, many more inner doubles will be 
resolved. SKA VLBI studies along with SKA-Mid will 
greatly help to unravel the structure of the steep spectrum and 
GPS cores of radio galaxies.

\subsection{The inner jet dynamics and ram pressure balance equation}

In the context of astrophysical fluid dynamics, the dynamics of jets of 
radio galaxies describes the behavior of high-velocity flows through an 
ambient medium. A critical aspect of understanding these dynamics is the 
ram pressure balance equation, which governs the interaction between the 
jet and its surrounding environment. Ram pressure arises from the momentum 
flux of a fluid moving at high velocity relative to an object. In the 
context of jets, this pressure is exerted by the ambient medium on the jet 
as it propagates through it. Understanding ram pressure is essential for 
analyzing the confinement and collimation of jets, as it directly influences 
their structure and stability \citep{Clarke&Carswell2007}. The ram pressure 
balance equation is a fundamental principle that describes the equilibrium 
between the jet's momentum flux and the opposing pressure from the 
ambient medium. In a steady-state scenario, the non-relativistic equation can 
be written as 
\begin{equation} 
{\dot M} v_j^2 = P_{ram} A_{h} , 
\label{eqn_rampress-NR}
\end{equation}
where ${\dot M}$ is the mass flow rate of the jet, $v_j$ is the velocity of 
the jet, and $A_h$ is the cross-sectional area of the jet head. This balance 
dictates the morphology and propagation speed of the jet, as it must counteract 
the external ram pressure to maintain its structure \citep{Castorena.etal2021}

%%-------------------------------------------------------------
%%-------------------------------------------------------------
Inner jets of double-double radio galaxies are ploughing through the relic 
radio plasma of outer lobes. Therefore, the inner lobe dynamics should be 
somewhat different from that of the outer lobes
if thermal matter does not get entrained into the outer relic lobes of
the episodic radio galaxies. Whether thermal matter has entrained into
the outer relic lobes has to be studied through measurements of
RM and/or depolarization (Section \ref{sec:polarization}). 
\cite{Kaiser.etal2000} used a ram pressure balance equation to
estimate the jet-head (i.e., hotspot) speed ($v_{\text{h}}$) through
the ambient medium. They used a semi-relativistic equation which is
\begin{equation}  
v_{\text{h}} =  \sqrt{ \frac{Q_j}{cA_h \rho^0_a} } 
\label{mombalance_Kaiser}
\end{equation}
where $Q_{\text{j}}$ is the jet power, $c$ is the speed of light,
$A_{\text{h}}$ is the jet head cross sectional area, and
$\rho^0_{\text{a}}$ is the rest mass density of the ambient medium.
With the application of the above
equation, \cite{Kaiser.etal2000} attempted to predict that there must
be thermal matter entrained into the relic lobes in order for 
hotspot formation to be possible. That is because too
low an ambient medium density would make the jet head move with a speed
close to the jet bulk speed. That would make the jet ballistic and no
hotspot would be observed. Since we see the hotspots in the inner jets
of double-double radio galaxies, according to \cite{Kaiser.etal2000},
there must be thermal matter entrained into the outer relic lobes.
However, \cite{Konar+Hardcastle13} showed that the ram pressure
balance equation should be made fully relativistic to generalize this conclusion. The
equation they derived is
\begin{equation} 
\beta_{\text{h}}= \left[ \frac{1}{1 + \sqrt{ \frac{\beta_{\text{j}} c
        A_{\text{h}} w_{\text{a}} }{Q_{\text{j}}}       }  }  \right]
\label{mombalance_KH}
\end{equation}
where $\beta_{\text{h}}$ and $\beta_{\text{j}}$ are the hotspot speed
and jet bulk speed in units of speed of light $c$, $A_{\text{h}}$ is
the jet-head cross section, $w_{\text{a}}$ is the enthalpy density of
the ambient medium, and $Q_{\text{j}}$ is the jet power.
Eq. \ref{mombalance_Kaiser} is a special case of
eq.\ \ref{mombalance_KH} in a non-relativistic situation, and hence
eq.\ \ref{mombalance_KH} is to be preferred, since we do not know
from first principles whether the dynamics are relativistic.
\cite{Konar+Hardcastle13} showed that even if there is no thermal
matter entrainment and the non-thermal lobe plasma consists of
electron-positron pair plasma (hereafter $e^{-} - e^{+}$ plasma) and
the number density of pair plasma particles is in the range
$10^{-9}-10^{-8}$ cm$^{-3}$, hotspot formation is highly plausible.
Therefore, the inner jet dynamics cannot tell us if thermal matter
entrainment into the outer relic lobes has taken place or not.
Instead, the measurement of RM internal to the lobes could tell us if
such entrainment has occurred. Then one can apply the relativistic ram
pressure balance equation from \cite{Konar+Hardcastle13} to study the
inner jet dynamics of episodic radio galaxies.

As discussed in Section \ref{sec:polarization}, Low may well
be able to help us conclude if there is significant thermal matter
entrainment into the outer relic lobes of episodic radio galaxies. If
entrainment happens and we are able to obtain some estimates of the
thermal matter density inside lobes, then that will be used to study
the inner jet dynamics of episodic radio galaxies.
\cite{Konar+Hardcastle13} used only non-thermal cocoon matter (no
thermal matter) as the ambient medium for the inner jet propagation to
estimate the jet-head speed and concluded that formation of hotspots
at the jet-head of inner jets is quite plausible, and the jet-head
speed has to be relativistic. However, this does not guarantee that
there is no thermal matter inside the lobes. There can always be some
amount of thermal matter which was entrained into the lobes. However,
the amount of thermal matter inside the lobes cannot be very high \citep[cf.][]{Croston+18} as
there is observational evidence for bow shock formation (see
Fig.-6 of \cite{Konar+Hardcastle13} around the jet-head of inner jets
propagating through the non-thermal cocoon matter of outer relic
lobes. As \cite{Konar+Hardcastle13} argue, this is possible only if there is a jet
termination shock at the inner jet-head and the jet head propagates
with relativistic speed which is faster than the magnetosonic speed of
the tenuous non-thermal ambient medium consisting of primarily
non-thermal cocoon matter of the outer lobes. Because of the
relativistic motion of the inner jet-heads, the radiation from the
compact components of the jet-heads (which are due to jet termination
shocks) undergoes relativistic boosting in the jet side and relativistic
de-boosting in the counter-jet side. That makes the jet head in the jet
side (which is moving towards us) appear as a prominent compact
component suppressing the bow shock structure. Further, the jet head
of the counter jet side loses the prominence of the compact component
in the radiation because of the relavistic de-boosting effect and the bow
shock becomes prominent. This, as pointed out by
\cite{Konar+Hardcastle13}, is a clear observational signature of the
relativistic motion of the inner hotspots. Again with the SKA, because
of very high resolution and sensitivity, for a large number of sources
we will be able to detect the observational signatures of relativistic
motion of the inner jet-heads. If, on the other hand, the inner jet-heads
of some episodic radio galaxies may show strong compact components in
the jet-heads of both jets and counter jets, this will mean that the
sufficient thermal matter entrainment into the relic outer lobes has
taken place in those episodic radio galaxies to mean that the inner
lobes no longer propagate at relativistic speeds. We will get a much
better picture with the use of SKA data which will allow us to resolve
the inner doubles.

\subsection{Injection index and jet power}

\cite{Konar+Hardcastle13} found that (i) the injection spectral
indices of inner and outer doubles are quite similar in values within
the error bars for a sample of 8 double-double radio galaxies, and
(ii) there exists a correlation between injection spectral index and
jet power for a sample of radio galaxies showing both single and
double episodes together. This finding (ii) has triggered rethinking
over the correlation between spectral index ($\alpha$) and radio power
at a given frequency ($P_{\nu}$), and the correlation between spectral
index ($\alpha$) and redshift ($z$) as published by
\cite{Laing&Peacock1980} for a flux density limited sample. Since in a
flux density limited sample the $z$ and $P_{\nu}$ are correlated,
$\alpha - P_{\nu}$ correlation can give rise to $\alpha - z$
correlation, and vice versa. Therefore, there has been a long-standing
debate on which of these is the primary correlation \citep[e.g.][]{Blundell+99}.
\cite{Konar+Hardcastle13} and more recently \cite{Pinjarkar.etal2025}
revisited this discussion from the perspective of their particular
observations. \cite{Konar+Hardcastle13} explained the similar
injection index in the two episodes of double-double radio galaxies by
introducing (a) strong shock at the hotspots of both inner and outer
doubles along with spine-sheath structure of the jets with the spine
moving with Lorentz factor $\Gamma_{\text{j}}>10$, (b) similar jet
power in two episodes with similar bulk Lorentz factor of the jet
flow. To explain the correlation between injection spectral index
($\alpha_{\text{inj}}$) and jet power ($Q_{\text{jet}}$), they
introduced the conjecture that higher jet power in sources is due to
higher energy density in the jet fluid and not due to higher jet
Lorentz factor. High energy density (likely to be in equipartition)
will have higher magnetic field which enhances the synchrotron loss to
steepen the originally produced injection index to a steeper value.
Thus, higher jet power (or, higher monochromatic luminosity) will
produce steeper observed injection spectral index. Thus
\cite{Konar+Hardcastle13} interpreted that the correlation between
spectral index (or, injection spectral index) and jet power is the
primary correlation and one between spectral index (or, injection
spectral index) and redshift is the secondary one. The latter
correlation arises in a flux density limited sample because in such a
sample there exists a correlation between redshift and radio power at
a given frequency (or, monochromatic luminosity). In a recent study,
\cite{Pinjarkar.etal2025} found the evidence that the correlation
between spectral index and luminosity is the primary correlation,
which is in the line of the conclusion drawn by
\cite{Konar+Hardcastle13} and supports original arguments by
\cite{Blundell+99}.

The above-mentioned results in (i) and (ii) give interesting
constraints on AGN jet dynamics. With the SKA, the emission of inner
and outer doubles will be clearly separated for many double-double
radio galaxies for a wide range of frequencies starting from 350 MHz
to 15 GHz in Mid. Verifying the correlation between the injection
spectral indices of inner and outer doubles of double-double radio
galaxies with a larger sample will be feasible with the advent of SKA.
This will firmly establish if jet powers remain the same in each
episode of episodic radio galaxies, or if this is a selection effect
based on our existing small samples. In turn, that will provide an
important constraint on models of the jet launching mechanism, jet
dynamics and the cause of episodic jet activity in radio loud AGN.

If detailed spectral ageing analysis is done for a large number of
such sources with high resolution and sensitivity, we can also verify
the correlation between injection spectral index and jet power. If
verified, then we can even firmly establish that the correlation
between spectral index and radio power is the primary correlation and
the correlation between spectral index and redshift can arise because
the radio power and spectral index are correlated in a flux limited
sample. Given that the SKA will be highly sensitive, we can detect a
large number of faint radio galaxies at higher redshifts over a broad
range of frequencies. Therefore, we will be able to construct a sample up to a
certain redshift such that all radio galaxies within that redshift
will be detected in an SKA survey. The sample then will not be
flux-density limited. We expect in such a situation that there will
not be any correlation between redshift and radio power. In such a
sample we must see the correlation between spectral index (or,
injection spectral index) and jet power. This correlation will appear
as the correlation between spectral index and radio power at given
frequency (i.e., monochromatic luminosity). Hence the correlation
between injection spectral index and jet power will be firmly
established as the primary correlation. This would be best done in a
deep field study like that of \cite{Pinjarkar.etal2025}, but with
observations at comparable depths in all SKA bands to avoid the
sensitivity-dependent biases in observable spectral index which
affected that work. For example, extending the MIGHTEE fields used by
\cite{Pinjarkar.etal2025} to give good uniform sensitivity from Low to
Mid band 5 frequencies could be done by the SKA in a few tens of
hours.

\subsection{Testable prediction}
\label{subsec:3.6}
\cite{Konar+Hardcastle13} found similar injection spectral indices
in two episodes in episodic radio galaxies. If this is due to similar
jet power, as is their interpretation, their work suggests that in the
population of radio galaxies, there is a significant fraction of
strong radio galaxies which have similar jet power in two consecutive
episodes. This in no way means that there are no sources with
dissimilar jet power in two episodes. With the discovery of new
sources with the SKA, we are hopeful to find a population of episodic
radio galaxies with dissimilar injection spectral indices which would
mean that some of those sources must have dissimilar jet powers in the two
episodes. This would then allow an investigation of the factors that
can change or preserve the jet power in different episodes. The 
underlying parameters responsible for preserving jet power in two
episodes might be different in radiatively efficient and inefficient
objects (LERGs and HERGs: Section \ref{sec:intro}) although both classes appear to be able to
produce episodic radio galaxies. We might get clues to those underlying
parameters, once detailed investigation of a statistically significant
number of sources can be done through spectral ageing analysis
combined with X-ray and optical studies as has been done by
\cite{Konar.etal2006}, \cite{Konar.etal2012}, \cite{Konar.etal2013},
\cite{Jamrozy.etal2007}, \cite{Godambe.etal2009},
\cite{Konar+Hardcastle13}, and \cite{Konar.etal2019}. From the work of
\cite{Konar+Hardcastle13} and \cite{Konar.etal2019}, we can think of a
few causes of dissimilar injection spectral indices in two episodes.
We discuss them here. The injection spectral indices can be different
in two episodes if (a) the black hole mass changes significantly
from one episode to the other through matter accretion via radiatively
efficient accretion in the radio quiet mode of AGN during a quiescent phase
of jet activity, (b) the black hole spin changes due to angular
momentum accumulation from the accreted matter, (c) the mode of
accretion changes from radiatively efficient to inefficient, or vice versa,
irrespective of any change in black hole mass and spin, (d) the new
jet activity is misaligned compared to the previous episode because of
some large scale perturbation which may cause disturbance to the other
underlying parameters deciding jet power, (e) the jet Lorentz factor
changes significantly, or (f) the jet
Lorentz factor is less than 10 so that the Mach number of the jet
termination shock drastically varies in two episodes because of the
diverse ambient medium density through which the jets propagate in two
different episodes.

There may not be direct observable signatures for all the cause of
dissimilar injection spectral indices in two episodes. However, one of
our testable predictions is that we are likely to find dissimilar
injection spectral indices in all or a large fraction of misaligned
episodic radio galaxies. The episodic radio galaxy 3C\,293
\citep{Konar+Hardcastle13} and the central radio galaxy of the cluster
RBS 797 \citep{Ubertosi.etal2024} are illustrative examples of
misaligned episodic radio galaxies with dissimilar injection spectral
indices in two different episodes of jet activity. If through detailed
spectral ageing analysis we can determine the spectral age and
estimate the total energy budget of the lobes with equipartition
assumption, then in principle we will be able to estimate the jet
power for both pairs of lobes in episodic radio galaxies. If jet
powers of inner and outer doubles in an episodic radio galaxy are the
same, and still the injection spectral indices are dissimilar for two
episodes; we can interpret the dissimilarity to be due to the Lorentz
factor of the jet spine having a value less than 10 or so (see
\cite{Konar+Hardcastle13}). If we measure highly dissimilar jet powers
then we would expect to get dissimilar injection spectral indices in
two episodes. Jet power differences can be due to various reasons which
cannot always be distinguished through observations. With the advent of SKA
with its high resolution and sensitivity, detailed spectral ageing
studies of both inner and outer lobes will be possible and the picture of
radio galaxy dynamics revealed by episodic sources will be much clearer than what it is today.

\subsection{Lower-power and spiral-host episodic sources}

As discussed above, most of our current understanding of jet duty
cycles arises from multi-frequency spectral-aging studies of episodic
radio galaxies hosted by elliptical galaxies. However, the phenomena
of episodic radio jets at smaller linear scales can also be seen in
nearby Seyfert galaxies. The galaxies Mrk 6, NGC2992 and NGC2639 have
three episodes of radio jets/lobes each on different scales and
different orientations to demonstrate their episodic nature
\citep{Rao.etal2023} and references therein). These episodes are not
as distinct as in episodic FRII radio galaxies. X-ray imaging studies
of NGC 5813 shows three pairs of cavities clearly suggesting episodic
non-thermal lobes pushing the thermal gas aside
\citep{Randall.etal2011}. A small but growing number of large radio
galaxies have been identified in spiral or disk hosts. Since the
earliest discoveries, the known sample of such systems has expanded to
roughly three dozen \citep{Ledlow.etal1998, Yuan.etal2024}. Among
these, three sources exhibit megaparsec-scale radio lobes with
evidence of episodic activity — Speca \citep{Hota.etal2011},
J23453268-0449256 \citep{Bagchi.etal2014} and LEDA 896325
\citep{Sethi.etal2025}. Detailed investigations of these rare systems
has remained limited, and it is yet to be established whether their
properties differ systematically from those of giant episodic radio
galaxies hosted by ellipticals. Given that the gaseous environments of
spiral and elliptical galaxies differ markedly, the propagation of
jets and expansion of radio lobes through these media are expected to
have distinct implications for lobe energetics and radiative
evolution. As the hosts of these radio galaxies appear to have evolved
primarily through passive processes rather than major mergers,
assembling a larger sample including high-redshift source will be
crucial for probing their cosmological evolution beyond $z = 2-3$ or
``cosmic noon'' \citep{Florez+20}. Future progress in this area will depend on
sensitive, broad-band and wide-field observations with the SKA to
identify additional disk-hosted episodic systems, complemented by
high-resolution optical imaging (e.g. with \textit{Euclid}) to confirm
their spiral/disk host morphologies. The SKA will provide the
sensitivity and resolution necessary to carry out spectral ageing
studies of these faint radio sources with episodic jets out to cosmological distances.

\section{Remnant radio galaxies, their identification and their
  characterization}
\label{sec:remnant}
\subsection{Introduction to remnant radio galaxies}

The jet activity of AGN in radio galaxies can persist for up to
hundreds of Myr, during which the radio source can grow to sizes of a few hundred kiloparsecs and, in some cases, extend to megaparsec scales \citep{Parma+07,Machalski+07,Pinjarkar+23}. Once this active phase ends, the central engine either ceases activity or falls below the threshold required to sustain powerful jets, causing the outflows to shut down and the radio lobes to begin fading \citep{Parma+99}. This fading stage, referred to as the {\em remnant} or {\em dying} phase, is characterized by the disappearance of the radio core and jets, while the lobes remain detectable for several tens of Myr before becoming undetectable due to cumulative radiative and dynamical losses \citep{Slee+01,Murgia+11,Lalremnant2021}. The remnant phase is generally thought to be much shorter than the preceding active phase, implying that remnant sources represent a brief terminal stage in the evolution of radio galaxies \citep{Morganti+17}. Consequently, remnant radio galaxies (hereafter remnants) are expected to be intrinsically rare, and their short lifetimes and low space densities mean that their physical properties and demographics remain poorly constrained.

Recent low-frequency studies have demonstrated that remnant radio emission can provide crucial constraints on the duty cycle of AGN jets. In particular, \citep{Pandey-Pommier2016} used LOFAR MSSS and TGSS observations to detect relic lobes associated with previous activity cycles in nearby radio galaxies, highlighting how faint, diffuse emission traces earlier jet episodes that have ceased. Their spectral analysis revealed a correlation between the mean particle age of the relic emission and the central AGN properties, indicating that low-frequency relics preserve a record of past activity and can be used to infer the relative durations of active and quiescent phases.
Such studies highlight the diagnostic potential of low-frequency observations for identifying dying or restarted radio galaxies. In addition to isolated radio galaxies, studies of brightest cluster galaxies in cool-core clusters reveal that recurrent jet activity plays a critical role in regulating the thermal state of the intracluster medium \citep{Hamer2016}. These observations indicate that AGN duty cycles are not only imprinted in relic lobes but also in
the cumulative feedback energy deposited into the surrounding environment, highlighting the environmental dependence of episodic AGN activity.

To investigate the properties of remnants, several studies have conducted systematic searches, primarily utilizing low-frequency ($<$1.4 GHz) radio surveys \citep[e.g.][]{Parma+07,Murgia+11,Brienza+17,Godfrey+17,Mahatma+18,Jurlin+20,Dutta+23}. Low-frequency observations are particularly well suited for identifying remnants, as the synchrotron spectra of aged lobe plasma steepen significantly at higher frequencies. Consequently, samples of Ultra-Steep Spectrum (USS) radio sources have often been employed as efficient tracers of remnant candidates. However, simulations have clearly demonstrated that low-frequency ($<1$ GHz) spectral selection is biased toward the oldest end of the remnant population and fails to identify younger remnants whose spectra have not yet steepened significantly \citep{Brienza+17,Godfrey+17}. \cite{Brienza+17} show that only a spectral selection that includes 5 GHz data --- where spectral steepening is stronger and occurs more rapidly --- can recover nearly the entire remnant population (up to 97\%). In the absence of this, complementary morphological selection criteria become paramount.

\cite{Mahatma+18} applied a morphological approach, selecting remnants based on the absence of a radio core in sensitive high-frequency observations. These approaches are complicated by the fact that remnant dynamics depends on both progenitor and environment properties \citep{Yates+23}. However, using morphological and spectral criteria, core non-detection is inherently limited by observational sensitivity, and weak cores associated with either fading or rejuvenated AGN activity may still be present \citep{Jurlin+21}. Distinguishing between dying and restarted cores remains challenging, and there is evidence that low-level AGN activity may persist in massive galaxies regardless of evolutionary stage \citep[see][]{Sabater+19}. \cite{Singh+21} and \cite{Dutta+23} employed both morphological and spectral criteria to assemble the largest sample of faint (8.0 mJy at 325 MHz) remnants identified to date, and this approach will need to be adopted in future SKA studies. 
The remnant fraction ($f_{\rm rem}$) is found to be only 5-8\% even in deep radio fields, with the exact value depending on the flux density threshold at which the sources are probed \citep{Jurlin+21,Lalremnant2021,Dutta+23}. Owing to their scarcity, the evolutionary pathways of radio galaxies during the remnant phase are not well understood, but if a selection strategy can be devised then the SKA should be able to find them in large numbers.

\citet{Shabala+20} used sensitive LOFAR samples of active and remnant radio galaxies from \citet{Jurlin+20} to place strong constraints on the distribution of jet lifetimes. Starting with a wide range of plausible progenitors with jet power and lifetime distributions constrained to be consistent with the observed active radio galaxy populations, these authors used a dynamical radio source model \citep{Turner+15} to predict the observable remnant fraction for each scenario. Their results showed that the overall population of radio galaxy jets has a pink noise-like power spectrum in age \citep{Shabala+20}; in other words, there are many more short-lived jets than long-lived ones. This requirement is placed by the relatively high ($\sim$10\%) observed remnant fraction \citep{Turner+15}, and the rapid fading of remnant lobes \citep{Turner+20}. The SKA will be able to extend such work to larger samples, and definitively test the hints that the high-power radio galaxy population follows a different duty cycle to lower-power counterparts \citep{Hardcastle+19}.

\subsection{A case study of a remnant source in the COSMOS deep field}
\begin{figure*}
    %\centering
    \includegraphics[scale=0.046]{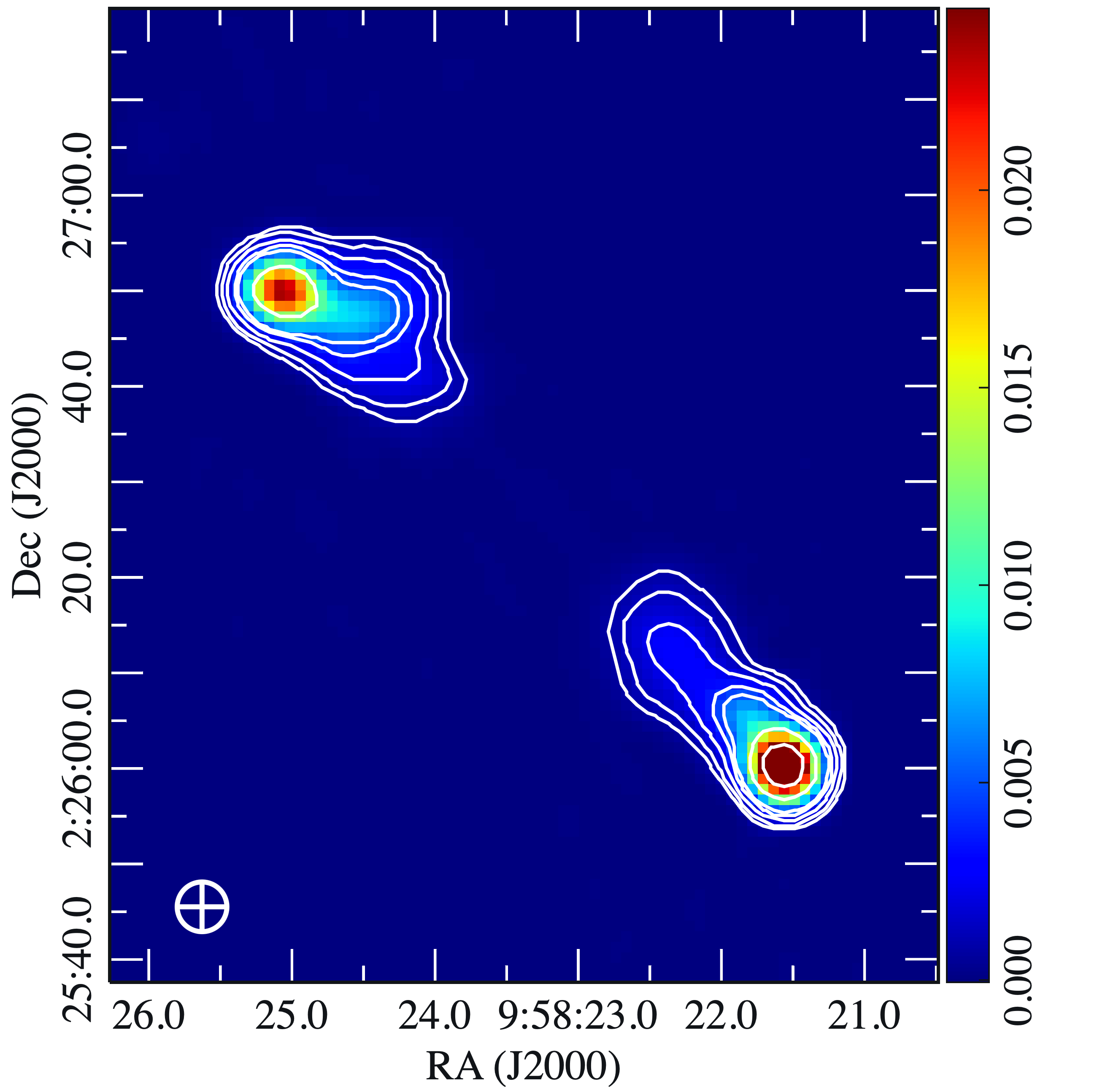}
    \includegraphics[scale=0.17]{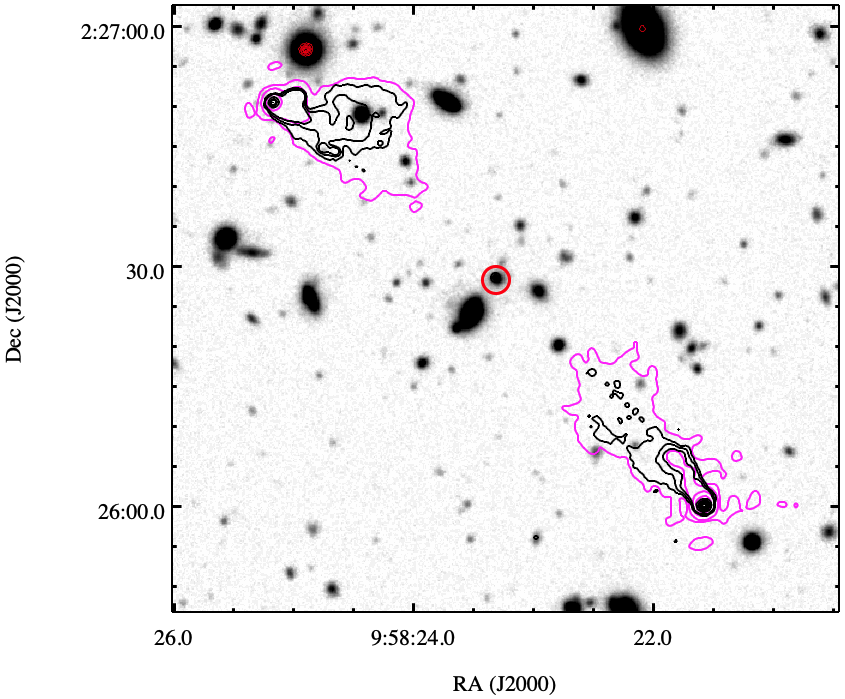}
    \includegraphics[scale=0.23]{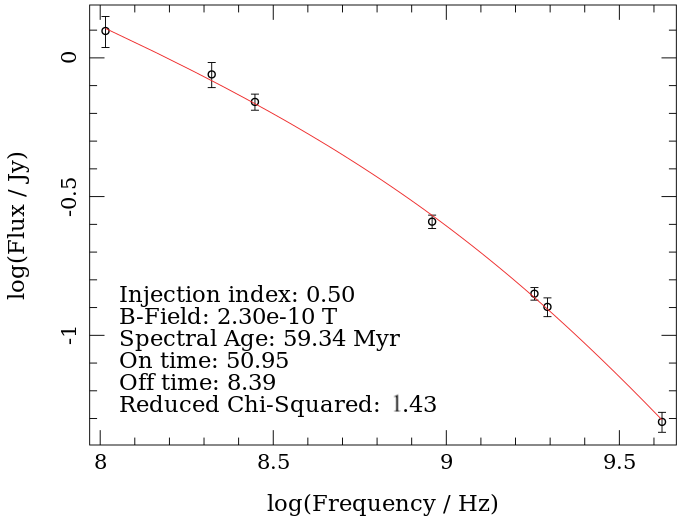}
    \caption{{\em Left panel:} MIGHTEE image of J095823$+$022628 in the COSMOS field with radio contours at 3$\sigma$ $\times$ (1, 2, 4, 8, 16, \dots). {\em Middle panel:} 1.4 GHz VLA (in black) and 3.0 GHz VLA (in magenta) radio contours overlaid on the Ultra-VISTA NIR image, with the red circle marking the potential optical host. {\em Right panel:} Best-fitted radio SED of the source.}
    \label{fig:ContourPlots}
\end{figure*}

J095823$+$022628 serves as an example of a remnant radio galaxy in the COSMOS field, representative of the population of aged radio sources that upcoming SKA surveys are expected to uncover in large numbers, enabling detailed studies of their spectral evolution and energetics across cosmic time. 
%We report the discovery of a new remnant radio galaxy, J095824$+$022624, in the COSMOS field. 
Morphologically, no radio core is detected in high-resolution imaging from MIGHTEE at 1284 MHz, superMIGHTEE \citep{Lal2025} at 650 MHz, FIRST at 1.4 GHz, or VLASS at 3.0 GHz. This source also exhibits hotspot-like features in both lobes (see left panel of Figure~\ref{fig:ContourPlots}), indicative of a relatively young remnant phase. The absence of a compact core in these datasets strongly supports its identification as a remnant. This source has also been examined by \cite{Vardoulaki+19,Vardoulaki+21}, who used the more sensitive 1.4 GHz \citep{Schinnerer+10} and 3.0 GHz VLA-COSMOS \citep{Smolcic+17} surveys to identify the radio core. They detected a faint core with a core prominence ($S_{\rm core}/S_{\rm total}$) of 2$\times$10$^{-3}$ (see middle panel of Figure~\ref{fig:ContourPlots}). In the literature, remnant radio galaxies are typically characterised by very low core prominence values, generally in the range 10$^{-4}$ to 3$\times$10$^{-3}$ \citep[see][]{Brienza+17}. The measured value for this source falls squarely within this range, supporting our interpretation that, morphologically, it satisfies the criteria for being a young remnant. In addition, the radio SED exhibits pronounced spectral curvature (see right panel of Figure~\ref{fig:ContourPlots}), providing further evidence that J095823$+$022628 is a genuine remnant source. The potential optical host galaxy of this source (see middle panel of Figure~\ref{fig:ContourPlots}) has a spectroscopic redshift of $z = 1.168\pm 0.004$ \citep[VIMOS;][]{Lilly+23}. The details of the spectral ageing analysis for this source are presented below.

Active radio galaxies typically exhibit a power-law injection radio
spectrum in the location where particles are being injected. Over time, continuous radiative losses in the lobe plasma cause the spectrum to deviate from a simple power law, producing a spectral break. Because high-energy electrons lose energy more rapidly than low-energy ones, this break occurs at high frequencies, beyond which the spectrum steepens relative to the injection index (${\alpha}_{\rm inj}$).

For active sources, the spectral shape is often reasonably well described by the continuous injection (CI) model \citep[CI$_{\rm ON}$;][]{Kardashev+62,Jaffe+Perola73}, which assumes a steady supply of relativistic plasma into the lobes over an active lifetime $t_{\rm active}$. In the remnant phase, once the jets switch off and no fresh particles are injected, the existing electron population continues to lose energy through radiative processes. This results in the formation of an additional spectral break at a higher frequency (${\nu}_{\rm b,~high}$) beyond the original injection break (${\nu}_{\rm b,~low}$). The lower break frequency, ${\nu}_{\rm b,~low}$, represents the spectral steepening accumulated during the active phase, whereas ${\nu}_{\rm b,~high}$ indicates the frequency above which the oldest electrons—those injected before the jet switched off—have largely expended their energy. The spectrum falls off exponentially beyond ${\nu}_{\rm b,~high}$, and as the source ages, both $\nu_{\rm b,~low}$ and ${\nu}_{\rm b,~high}$ drift toward lower frequencies. The ratio of the two break frequencies encodes the relative duration of the remnant phase, expressed as 
\begin{equation} 
\label{eq:1}
\frac{t_{\rm quies}}{t_{\rm s}} = \left(\frac{\nu_{\rm b,~low}}{\nu_{\rm b,~high}}\right)^{0.5},
\end{equation}
where $t_{\rm s} = t_{\rm active} + t_{\rm quies}$ is the total source age. For sufficiently old remnants, ${\nu}_{\rm b,~high}$ approaches ${\nu}_{\rm b,~low}$. The relation between the spectral age $t_{\rm s}$ and the observed break frequency ${\nu}_{\rm b}$ is given by \citep[see][]{Komissarov+94,Slee+01,Parma+07}:
\begin{equation} 
\label{eq:2}
t_{\rm s} = 1590 \left[\frac{B_{\rm eq}^{0.5}}{(B_{\rm eq}^2 + B_{\rm CMB}^2) \sqrt{\nu_{\rm b} (1 + z)}}\right]~{\rm Myr,}
\end{equation}
where $B_{\rm eq}$ is the equipartition magnetic field, $B_{\rm CMB} = 3.25(1+z)^2$ is the inverse Compton equivalent magnetic field in units of $\mu$G and $\nu_{\rm b}$ is in GHz. This formulation accounts for radiative losses but neglects adiabatic expansion \citep{English+19,Turner+20}.

Spectral age estimates also require knowledge of the magnetic field strength. For this purpose, we adopted the equipartition magnetic field calculated using the {\sc PySynch}\footnote{\url{https://github.com/mhardcastle/pysynch}} code \citep{Hardcastle+98}, and assumed magnetic field strength, $B = 0.4B_{\rm eq}$ \citep[see][]{Croston+05,Ineson+17,Mahatma+19}. For this source, the magnetic field strength is estimated to be 2.3 $\mu$G ($B_{\rm eq}$ = 5.72 $\mu$G). In our modeling, we also used a constant injection index of $\alpha_{\rm inj} = 0.5$, consistent with expectations for synchrotron emission from a power-law distribution of relativistic electrons.

To model the radio SED of our remnant source, we adopt the continuous injection-off model \citep[CI$_{\rm OFF}$ or KGJP;][]{Komissarov+94}, which explicitly incorporates the remnant phase duration $t_{\rm quies}$ following the cessation of jet activity after an active phase of length $t_{\rm active}$. We constructed the radio SED using flux density measurements from 74 MHz VLSSr, 150 MHz TGSS, 200 MHz GLEAM, 650 MHz superMIGHTEE, 1284 MHz MIGHTEE, 1.4 GHz NVSS, and 3.0 GHz VLASS surveys. The parameters obtained using CI$_{\rm OFF}$ are given in Table~\ref{tab:table1} and the best-fitted radio SED is shown in the right panel of Figure~\ref{fig:ContourPlots}. The total source age ($t_{\rm s}$) is estimated to be 59.34$^{+5.13}_{-5.32}$ Myr, with an active phase duration ($t_{\rm active}$) of 50.95$^{+4.97}_{-5.12}$ Myr and a remnant phase duration ($t_{\rm quies}$) of 8.39$^{+1.26}_{-1.45}$ Myr. The ratio $t_{\rm quies}$/$t_{\rm s}$ further suggests that the source has spent only 14\% of its lifetime in the remnant phase, consistent with its classification as a young remnant. Because the radio lobes fade on short timescales after the jets stop injecting energy, all observational searches are strongly biased toward detecting the youngest remnant sources \citep{English+19,Shabala+24}.   

\begin{table*}
  \caption{Spectral ageing parameters for J095823$+$022628 obtained by using CI$_{\rm OFF}$ model.}

	\begin{minipage}{180mm}
		%\centering
        \tabcolsep=1mm
       \small
		\label{tab:table1}
		\begin{tabular}{ccccccccccc}
			\hline
			Source & Redshift & $\alpha_{\rm inj}$ & B & $\nu_{\rm b,low}$ & $\nu_{\rm b,high}$ & t$_{\rm s}$ & t$_{\rm active}$ & t$_{\rm quies}$ & t$_{\rm quies}$/t$_{\rm s}$  \\
			Name & ($z$) & & ($\mu$G) & (MHz) & (GHz) & (Myr) & (Myr) & (Myr) & \\
			\hline
			J095823$+$022628 & 1.168$\pm$0.004 & 0.50 & 2.3 & 564.97 & 5.38 & 59.34$^{+5.13}_{-5.32}$ & 50.95$^{+4.97}_{-5.12}$ & 8.39$^{+1.26}_{-1.45}$ & 0.14$^{+0.03}_{-0.03}$ \\
			\hline
		\end{tabular}
	\end{minipage}
\end{table*}

\subsection{Predictions with the SKA}
As discussed in subsection~\ref{subsec:3.3} and \ref{subsec:3.6}, the SKA will revolutionize the detection and characterization of remnants. Its unprecedented sensitivity and resolution are expected to uncover a previously unexplored population of faint remnants residing at high redshifts. With Low covering the 50-350 MHz range and Mid spanning 0.35-15.4 GHz, these facilities will provide unprecedented spectral coverage for detailed modeling of aged electron populations in radio galaxies. Mid band 1 is expected to achieve an rms sensitivity of $\sim 2\ \mu$Jy beam$^{-1}$ within one hour of integration (centered at 797.5 MHz), while Low will reach $\sim 12\ \mu$Jy beam$^{-1}$ at 200 MHz in the same time. The combination of large-area, deep, and multi-frequency surveys from the SKA and its pathfinders will thus enable the discovery of a statistically significant population of remnant radio galaxies spanning a wide range of evolutionary stages, providing transformative insights into the duty cycles and life histories of radio-loud AGN.
\par
Using the MeerKAT $L-$band array with a 139.6 h observation in the COSMOS field, we achieved an rms sensitivity reaching down to $\sim 2\ \mu$Jy beam$^{-1}$ and an angular resolution of $\sim6.0$ arcsec. In contrast, the SKA will deliver an order-of-magnitude improvement in both sensitivity and survey speed, reaching sub-$\mu$Jy levels with arcsecond-scale resolution in a fraction of the time. Such performance will revolutionize the study of remnant radio galaxies. Compared to current SKA pathfinders such as MeerKAT, uGMRT, LOFAR, and ASKAP, Mid is expected to increase the number of detectable remnant candidates by roughly a factor of 3-30 (point-source flux limit $S_{\rm lim,SKA} = S_{\rm lim,Old}/f$ and naive Euclidean source counts give $N \propto S^{-1.5}$, so $N_{\rm SKA}/N_{\rm Old} \approx f^{1.5}$), while Low could enhance detections by up to an order of 10-100, owing to its superior surface-brightness sensitivity to steep-spectrum, low-frequency emission. This improvement will enable comprehensive population studies of faint and extended remnants, providing unprecedented constraints on AGN duty cycles, spectral ageing, and the late evolutionary stages of radio galaxies.

\section{Summary}

The SKA will provide an unparalleled step forward in our understanding of radio galaxies, principally by providing very large samples of detailed images of objects that have hitherto been studied in such detail in very small samples or as single objects. We have emphasised in particular the resolution, sensitivity, and broad-band spectral capabilities of the Mid array in the possible projects that we have highlighted above, while not forgetting the contribution of Low to
observations of steep-spectrum extended emission and low-frequency polarization/Faraday rotation studies. Much of the work described above can be done comensally with high-frequency surveys, together potentially with pointed large-sample studies for high-frequency work. All of it is well within the capabilities of the SKA AA4 array considered here. We can confidently expect a transformational improvement in our understanding of the radio galaxy population by the end of the coming decade.

\bibliographystyle{abbrvnat-maxbibnames4.bst}
\bibliography{chapter}

\end{document}

%% file: journal-names.tex
\newcommand{\actaa}{Acta Astron.} % Acta Astronomica
\newcommand{\araa}{Annu. Rev. Astron. Astrophys.} % Annual Review of
                                % Astron and Astrophys
\newcommand{\aar}{Astron. Astrophys. Rev.} % Astrononmy and
                                % Astrophysics Review
\newcommand{\ab}{Astrobiol.} % Astrobiology
\newcommand{\aj}{Astron. J.} % Astronomical Journal
\newcommand{\apj}{Astrophys. J.} % Astrophysical Journal
\newcommand{\apjl}{Astrophys. J. Lett.} % Astrophysical Journal, Letters
\newcommand{\apjs}{Astrophys. J. Suppl. Ser.} % Astrophysical Journal,
                                % Supplement
\newcommand{\ao}{Appl. Opt.} % Applied Optics
\newcommand{\apss}{Astrophys. Space Sci.} % Astrophysics and Space Science
\newcommand{\aap}{Astron. Astrophys.} % Astronomy and Astrophysics
\newcommand{\aapr}{Astron. Astrophys. Rev.} % Astronomy and
                                % Astrophysics Reviews
\newcommand{\aaps}{Astron. Astrophys. Suppl.} % Astronomy and
                                % Astrophysics, Supplement
\newcommand{\baas}{Bull. Am. Astron. Soc.} % Bulletin of the AAS
\newcommand{\caa}{Chinese Astron. Astrophys.} % Chinese Astronomy and Astrophysics
\newcommand{\cjaa}{Chinese J. Astron. Astrophys.} % Chinese Journal of
                                % Astronomy and Astrophysics
\newcommand{\cqg}{Class. Quantum Gravity} % Classical and Quantum Gravity
\newcommand{\gal}{Galaxies} % Galaxies
\newcommand{\gca}{Geochim. Cosmochim. Acta} % Geochimica Cosmochimica Acta
\newcommand{\icarus}{Icarus} % Icarus
\newcommand{\jcap}{J. Cosmol. Astropart. Phys.} % Journal of Cosmology
                                % and Astroparticle Physics
\newcommand{\jgr}{J. Geophys. Res.} % Journal of Geophysics Research
\newcommand{\jgrp}{J. Geophys. Res.: Planets} % Journal of Geophysics Research: Planets
\newcommand{\jqsrt}{J. Quant. Spectrosc. Radiat. Transf.} % Journal of
                                % Quantitiative Spectroscopy and Radiative Transfer
\newcommand{\memsai}{Mem. Soc. Astron. Italiana} % Mem. Societa
                                % Astronomica Italiana
\newcommand{\mnras}{Mon. Not. R. Astron. Soc.} % Monthly Notices of the RAS
\newcommand{\nat}{Nature} % Nature
\newcommand{\nastro}{Nat. Astron.} % Nature Astronomy
\newcommand{\ncomms}{Nat. Commun.} % Nature Communications
\newcommand{\nphys}{Nat. Phys.} % Nature Physics
\newcommand{\na}{New Astron.} % New Astronomy
\newcommand{\nar}{New Astron. Rev.} % New Astronomy Review
\newcommand{\physrep}{Phys. Rep.} % Physics Reports
\newcommand{\pra}{Phys. Rev. A} % Physical Review A: General Physics
\newcommand{\prb}{Phys. Rev. B} % Physical Review B: Solid State
\newcommand{\prc}{Phys. Rev. C} % Physical Review C
\newcommand{\prd}{Phys. Rev. D} % Physical Review D
\newcommand{\pre}{Phys. Rev. E} % Physical Review E
\newcommand{\prl}{Phys. Rev. Lett.} % Physical Review Letters
\newcommand{\psj}{Planet. Sci. J.} % Planetary Science Journal
\newcommand{\planss}{Planet. Space Sci.} % Planetary Space Science
\newcommand{\pnas}{Proc. Natl Acad. Sci. USA} % Proceedings of the US
%National Academy of Sciences
\newcommand{\procspie}{Proc. SPIE} % Proceedings of the SPIE
\newcommand{\pasa}{Publ. Astron. Soc. Aust.} % Publications of the Astron.
%Soc. of Australia
\newcommand{\pasj}{Publ. Astron. Soc. Jpn} % Publications of the Astron.
%Soc. of Japan (note no full stop following Jpn)
\newcommand{\pasp}{Publ. Astron. Soc. Pac.} % Publications of the Astron.
%Soc. of the Pacific
\newcommand{\rmxaa}{Rev. Mexicana Astron. Astrofis.} % Revista Mexicana de
%Astronomia y Astrofisica
\newcommand{\sci}{Science} % Science
\newcommand{\sciadv}{Sci. Adv.} % Science Advances
\newcommand{\solphys}{Sol. Phys.} % Solar Physics
\newcommand{\sovast}{Soviet Ast.} % Soviet Astronomy
\newcommand{\ssr}{Space Sci. Rev.} % Space Science Reviews
\newcommand{\uni}{Universe} % Universe